\documentclass[12pt]{iopart}
\eqnobysec
\usepackage{amssymb}
\usepackage{epsf}

\def\be{\begin{equation}}
\def\ee{\end{equation}}
\def\bea{\begin{eqnarray}}
\def\eea{\end{eqnarray}}

\begin{document}
\jl{1}

\title[Random walk with memory]{On a random walk with memory and its relation to Markovian processes}

\author{Lo\"\i c Turban}

\address{Groupe de Physique Statistique, D\'epartement Physique de la Mati\`ere et des Mat\'eriaux,
Institut Jean Lamour\footnote[2]{Laboratoire associ\'e au CNRS UMR 7198.}, CNRS---Nancy Universit\'e---UPV Metz,\\
BP 70239, F-54506 Vand\oe uvre l\`es Nancy Cedex, France}
\ead{turban@lpm.u-nancy.fr}

\begin{abstract}
We study a one-dimensional random walk with memory in which the step lengths to the left and to the right evolve at each step in order to reduce the wandering of the walker. The feedback is quite efficient and lead to a non-diffusive walk. The time evolution of the displacement is given by an equivalent Markovian dynamical process. The probability density for the position of the walker is the same at any time as for a random walk with shrinking steps, although the two-time correlation functions are quite different.
\end{abstract}


\submitto{\JPA}

\section{Introduction} 
Introducing long-range correlations into a random walk may lead to drastic changes in its asymptotic behaviour. Depending on some parameter governing the strength and sign of the correlations, the Hurst exponent $\alpha$ of the mean-square displacement, given by $t^{2\alpha}$ at time $t$, can be modified. The dynamics can evolve from  diffusive ($\alpha=1/2$) to subdiffusive ($\alpha<1/2$), superdiffusive or persistent ($\alpha>1/2$). Such random walks with long-range memory have been extensiveley studied in the last years 
\cite{dickman03,hod04,schutz04,keshet05,paraan06,dasilva06,cressoni07,kenkre07,harris09}.

In this work we consider a one-dimensional random walk in which the walker tries to controll his wandering. In order to do so, at time $t$ the walker reduces the step length in the direction of the previous step at time $t-1$ while the sum of the step lengths in the two directions remains constant. For example, after a step to the right the length of a step to the right is reduced and the length of a step to the left is increased accordingly. The step lengths evolve in time in a non-systematic way and their actual values depend on the whole history. Thus the walk has a memory and belongs to the class of non-Markovian stochastic processes \cite{markov}.

We show that such a controlling process is quite efficient since the mean-square displacement saturates at long time, leading to a walk which is non-diffusive ($\alpha=0$). We prove its equivalence to a well studied Markovian dynamical process \cite{diaconis99}. The probability density for the position of the walker at time $t$ is found to be the same as for the random walk with shrinking steps \cite{delatorre00,krapivsky04,rador06a,rador06b,benjamini09a,serino10} although the two-time correlation functions for the position of the walker are quite different in the two walks. The singularities of the associated probability density have been much studied since the 1930s in the mathematics literature where the problem is known under the name of Bernoulli convolutions \cite{jessen35,kershner35,erdos39,erdos40,garsia62,solomyak95,peres00,benjamini09b}. 

The random walk with variable step lengths has found some physical applications. For example it has be used to expain some aspects of the motion of a Brownian particle in a shear flow \cite{bennaim92} as well as the spectral line broadening for single-molecule spectroscopy in a disordered solid \cite{barkai99,barkai00}.

The outline of the paper is the following: The details of the model are presented in section 2. The moment-generating function of the walk, which is actually given by the canonical partition function of a non-interacting Ising model, is obtained in section 3. In section 4 we show that the random walk with memory has the same time evolution as a Markovian dynamical process and that it has the same probability density as the random walk with systematically shrinking steps. We also compare the two-time correlation functions. We end with the conclusion in section 5.

\section{Model} 
The walker performs a discrete time one-dimensional random walk starting at the origin $X_0=0$ at $t=0$. One associates an Ising variable $\sigma_i=+1$ ($-1$) with the $i$th step to the right (left). The Ising variables  $\sigma_i$ $(i=1,t)$, taking on the values $\sigma_i=\pm1$ with equal probability, give a complete description of the walk. The step lengths at time $t$, $\ell_t^+$ for a right step and $\ell_t^-$ for a left step, satisfy the constraint 
\be
\ell_t^++\ell_t^-=2\,,
\label{constraint}
\ee
and evolve according to the following rules for $t>1$
\be\fl
\sigma_{t-1}=+1\longrightarrow
\left\{
\begin{array}{l}
\ell_t^+=\lambda\,\ell_{t-1}^+\\
\ell_t^-=2-\lambda\,\ell_{t-1}^+
\end{array}
\right.\,,
\qquad
\sigma_{t-1}=-1\longrightarrow
\left\{
\begin{array}{l}
\ell_t^-=\lambda\,\ell_{t-1}^-\\
\ell_t^+=2-\lambda\,\ell_{t-1}^-
\end{array}
\right.\,.
\label{def-1}
\ee
with $\ell_1^+=\ell_1^-=1$ and $0\leq\lambda\leq1$. The limit $\lambda=1$ corresponds to constant step lengths, i.e., to the Bernoulli random walk. When $\lambda<1$ the reduction of the step length in the direction of the previous step and its concomitant increase in the opposite direction reduce the wandering. When $\lambda=0$ the walker is restricted to stay for some random waiting time either at $X=+1$ or $X=-1$.

Due to the constraint \eref{constraint} a single step length is really needed to describe the evolution.
Let $\ell_t=\ell_t^+$, for $t>1$ we have
\be\fl
\sigma_{t-1}=+1\longrightarrow\ell_t=\lambda\,\ell_{t-1}\,,\qquad
\sigma_{t-1}=-1\longrightarrow\ell_t=2(1-\lambda)+\lambda\,\ell_{t-1}\,.
\label{def-2}
\ee
%%%%%%%%%%%%%%%%%%%%%%%%% fig 1 %%%%%%%%%%%%%%%%%%%%%%%%
\begin{figure} [tbh]
%\vspace{0.2cm}
\epsfxsize=14cm
%\begin{center}
\hskip 16mm
\mbox{\epsfbox{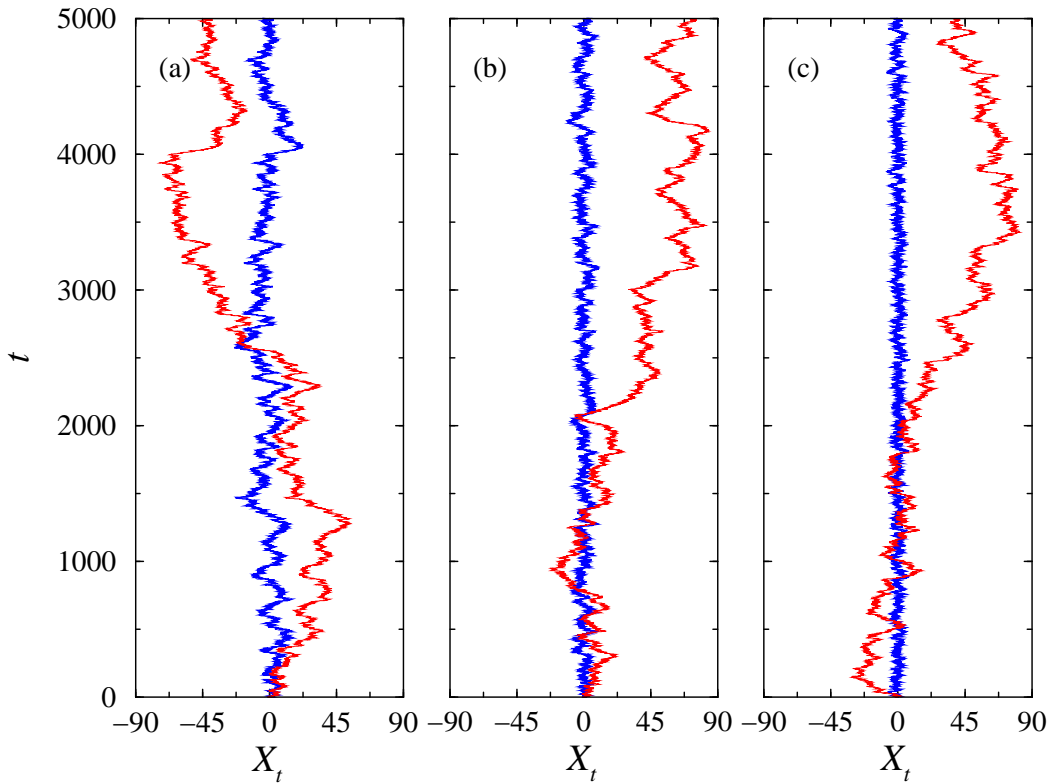}}
%\end{center} 
\vskip -.2cm
\caption{Random walks with memory (thick blue line) and Bernoulli random walks (thin red line) generated with the same random numbers for $\lambda=0.99$ (a), 0.95 (b), 0.9 (c). There is a strong reduction of the wandering for the walk with memory, even for values of $\lambda$ close to the Bernoulli value $\lambda=1$.}
\label{fig1-rw4}  \vskip 0cm
\end{figure}
%%%%%%%%%%%%%%%%%%%%%%%%%%%%%%%%%%%%%%%%%%%%%%%%%%%%
Making use of the Ising variables, equation~\eref{def-2} can be simply written as
\be
\ell_t=\lambda\,\ell_{t-1}+(1-\lambda)(1-\sigma_{t-1})\,,\qquad \ell_1=1\,.
\label{def-3}
\ee
Let $X_t$ be the position of the walker at time $t$ and $x_t$ its increment. One has
\be
X_t=X_{t-1}+x_t\,\qquad X_0=0\,,\qquad
x_t=\left\{
\begin{array}{cl}
\ell_t& \mbox{if $\sigma_t=+1$}\\
-(2-\ell_t)\ &\mbox{if $\sigma_t=-1$}
\end{array}
\right.\,,
\label{incr-1}
\ee
which translates into
\be
x_t=\ell_t+\sigma_t-1\,,\qquad x_1=\sigma_1\,.
\label{incr-2}
\ee
It is easy to verify that equations \eref{def-3} and \eref{incr-2} lead to the mean values
\be
\langle \ell_t \rangle=1\,,\qquad \langle X_t\rangle=\langle x_t\rangle=0\,,
\label{mv-1}
\ee
as expected for a symmetric walk.
In figure~\ref{fig1-rw4} random walks with memory are compared, for different values of $\lambda$, to Bernouilli random walks generated with the same random numbers. The feedback is quite efficient in controlling the wandering, even for $\lambda$ close to 1, the Bernoulli limit.

\section{Moment-generating function}
The moment-generating function for the position of the walker at time $t$ is given by
\be
P_\lambda(z,t)=\left\langle\exp(zX_t)\right\rangle=\frac{1}{2^t}\Tr_{\{\sigma\}}\exp(zX_t)
\label{pXt-1}
\ee
since the walks with $t$ steps are equiprobable with probability $1/2^t$. The trace over the Ising variables gives the sum over all the possible histories. Thus the moment-generating function is given by the canonical partition function at inverse temperature $z$ of an Ising model with Hamiltonian $-X_t(\{\sigma\})$.

In order to find out the expression of $X_t$ we first calculate $\ell_t$ by induction. At $t=2$ equation~\eref{def-3} gives
\be
\ell_2=\lambda+(1-\lambda)(1-\sigma_1)\,.
\label{lt-1}
\ee
Let us assume that the step length $\ell_t$ is given by
\be
\ell_t=\lambda^{t-1}+(1-\lambda)\sum_{j=1}^{t-1}\lambda^{t-j-1}(1-\sigma_j)\,,
\label{lt-2}
\ee
which agrees with \eref{lt-1} for $t=2$. Using~\eref{def-3}, one obtains
\be\fl
\ell_{t+1}\!=\!\lambda^t \!+\!(1\!-\!\lambda)\sum_{j=1}^{t-1}\lambda^{t-j}(1\!-\!\sigma_j)+(1\!-\!\lambda)(1\!-\!\sigma_t)
\!=\!\lambda^t +(1\!-\!\lambda)\sum_{j=1}^t\lambda^{t-j}(1\!-\!\sigma_j)\,,
\label{lt-3}
\ee
in agreement with \eref{lt-2}. Thus \eref{lt-2}, which remains true at $t+1$, is true at any time $t>1$.
Since
\be
(1-\lambda)\sum_{j=1}^{t-1}\lambda^{t-j-1}=(1-\lambda)\sum_{j=0}^{t-2}\lambda^j=1-\lambda^{t-1}\,,
\label{lt-4}
\ee
equation \eref{lt-2} reduces to
\be
\ell_t=1-(1-\lambda)\sum_{j=1}^{t-1}\lambda^{t-j-1}\sigma_j\,.
\label{lt-5}
\ee
The step length to the left, $\ell_t^-$, is related to the step length to the right, $\ell_t^+=\ell_t$, by
$\ell_t^-(\{\sigma\})=2-\ell_t=\ell_t^+(\{-\sigma\})$.

According to~\eref{incr-2}
\be
x_t=\sigma_t-(1-\lambda)\sum_{j=1}^{t-1}\lambda^{t-j-1}\sigma_j\,,
\label{xt}
\ee
and
\bea
\fl
X_t&=&\sum_{i=1}^t\sigma_i-(1-\lambda)\sum_{i=2}^t\sum_{j=1}^{i-1}\lambda^{i-j-1}\sigma_j
=\sum_{i=1}^t\sigma_i-(1-\lambda)\sum_{j=1}^{t-1}\sigma_j\sum_{i=j+1}^t\lambda^{i-j-1}\nonumber\\
\fl
&=&\sum_{i=1}^t\sigma_i-(1-\lambda)\sum_{j=1}^{t-1}\sigma_j\sum_{k=0}^{t-j-1}\lambda^k
=\sum_{i=1}^t\sigma_i-\sum_{j=1}^{t-1}(1-\lambda^{t-j})\sigma_j
=\sigma_t+\sum_{j=1}^{t-1}\lambda^{t-j}\sigma_j\nonumber\\
\fl
&=&\sum_{j=1}^t\lambda^{t-j}\sigma_j\,.
\label{Xt}
\eea
One may notice the close connection between $X_t$ and the deviation of $\ell_t$ from its mean value, 
$\ell_t-1=-(1-\lambda)X_{t-1}$, which follows from \eref{lt-5} and \eref{Xt}. Thus the scaled and centered random 
variable $(\ell_t-1)/(1-\lambda)$ has the same symmetric probability density as $X_{t-1}$. According to \eref{constraint} and \eref{def-1} one has $0\le\ell_t\le2$ so that $-1/(1-\lambda)\le X_t\le1/(1-\lambda)$ as expected.

%%%%%%%%%%%%%%%%%%%%%%%%% fig 2 %%%%%%%%%%%%%%%%%%%%%%%%
\begin{figure} [tbh]
%\vspace{0.2cm}
\epsfxsize=10cm
%\begin{center}
\hskip 16mm
\mbox{\epsfbox{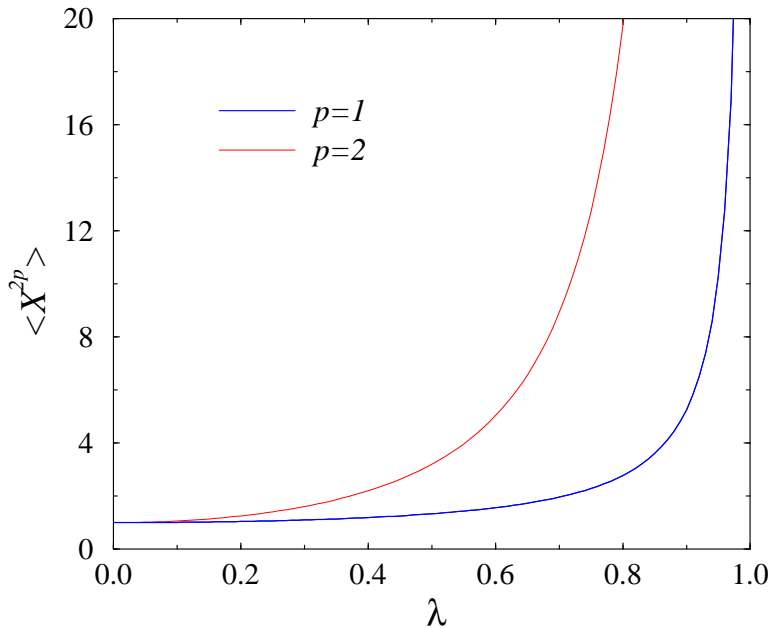}}
%\end{center} 
\vskip -.2cm
\caption{Moments of the asymptotic position of the walker $X=X_{t\to\infty}$ as a function of the control parameter $\lambda$.}
\label{fig2-rw4}  \vskip 0cm
\end{figure}
%%%%%%%%%%%%%%%%%%%%%%%%%%%%%%%%%%%%%%%%%%%%%%%%%%%%

The position of the walker at time $t$ is associated with a non-interacting Ising Hamiltonian. The moment-generating (or partition)
function follows from equations \eref{pXt-1} and \eref{Xt} and reads
\be\fl 
P_\lambda(z,t)=\frac{1}{2^t}\Tr_{\{\sigma\}}\exp\left[z\left(\sum_{j=1}^t\lambda^{t-j}\sigma_j\right)\right]
=\prod_{j=1}^t\cosh\left(z\lambda^{t-j}\right)=\prod_{j=0}^{t-1}\cosh\left(z\lambda^j\right)\,.
\label{pXt-2}
\ee
The moments of the position of the walker at time $t$ are given by:
\be
\langle X_t^n\rangle=\left.\frac{\partial^n P_\lambda}{\partial z^n}\right|_{z=0}\,.
\label{mom-1}
\ee
Since the walk is symmetric $P_\lambda(z,t)$ is an even function of $z$ and odd moments vanish. To calculate even moments, one expands $\cosh(z\lambda^j)$ in \eref{pXt-2} and collects terms with the same power of $z$ in the product:
\be\fl
P_\lambda(z,t)=\prod_{j=0}^{t-1}\left[\sum_{k=0}^\infty \frac{\lambda^{2jk}}{(2k)!}\,z^{2k}\right]
=1+\sum_{j=0}^{t-1}\frac{\lambda^{2j}}{2!}\,z^2+\left[\sum_{j=0}^{t-1}\frac{\lambda^{4j}}{4!}+
\!\!\!\!\!\sum_{0\leq j<k\leq t-1}\!\!\!\frac{\lambda^{2j}\lambda^{2k}}{(2!)^2}\right]z^4+\cdots
\label{pXt-3}
\ee
Applying \eref{mom-1} one obtains:
\be\fl
\langle X_t^2\rangle=\sum_{j=0}^{t-1}\lambda^{2j}=\frac{1\!-\lambda^{2t}}{1\!-\lambda^2}\,,\quad
\langle X_t^4\rangle=\sum_{j=0}^{t-1}\lambda^{4j}+6\!\!\!\!\!\sum_{0\leq j<k\leq t-1}\!\!\!\!\!\!\!\lambda^{2j}\lambda^{2k}=3\,\langle X_t^2\rangle^2-2\,\frac{1\!-\lambda^{4t}}{1\!-\lambda^4}\,.
\label{mom-2}
\ee
Figure \ref{fig2-rw4} shows the evolution with $\lambda$ of the values of $\langle X^2\rangle$ and $\langle X^4\rangle$ where $X=X_{t\to\infty}$ is the asymptotic position of the walker. The mean-square displacement remains quite close to 1 for values of $\lambda$ up to 0.7--0.8.

\section{Relation to Markovian processes}
The random variable $X_t$ can be generated by a {\em Markovian dynamical process} which follows from \eref{Xt}:
\be
X_{t+1}=\sum_{j=1}^{t+1}\lambda^{t-j+1}\sigma_j=\sigma_{t+1}+\lambda\sum_{j=1}^t\lambda^{t-j}\sigma_j
=\lambda X_t+\sigma_{t+1}\,.
\label{dp}
\ee
Such random iteration processes have been extensively studied in the mathematics litterature (see \cite{diaconis99} for a review). Equation \eref{dp} leads to a recursion relation for the moments
\be
\langle X_{t+1}^{2p}\rangle=\sum_{n=0}^{2p}{2p\choose n}\lambda^n\langle X_t^n\rangle
\langle\sigma_{t+1}^{2p-n}\rangle=\sum_{k=0}^p{2p\choose 2k}\lambda^{2k}\langle X_t^{2k}\rangle
\label{mom-3}
\ee
since $X_t$ and $\sigma_{t+1}$ are independant random variables and $\langle\sigma_{t+1}^{2p-n}\rangle=1$ when $n$ is even and otherwise vanishes. When $t\to\infty$ one obtains
\be
\langle X^{2p}\rangle=\frac{1}{1-\lambda^{2p}}\sum_{k=0}^{p-1}{2p\choose 2k}\lambda^{2k}\langle X^{2k}\rangle\,.
\label{mom-4}
\ee

The non-Markovian random walk with memory is also related to a Markovian one-dimensional random walk with shrinking steps \cite{krapivsky04}. In this model the position of the walker at time $t$ is given by
\be
X_t=\sum_{j=1}^t\ell_j\,\sigma_j=\sum_{j=1}^t\lambda^{j-1}\,\sigma_j\,,
\label{def-4}
\ee
where $\sigma_j$ takes on the values $\pm1$ with equal probability. There is now a {\em systematic reduction of the step length} which takes the same value for both directions. The moment-generating function is given by:
\be\fl 
P_\lambda(z,t)=\frac{1}{2^t}\Tr_{\{\sigma\}}\exp\left[z\left(\sum_{j=1}^t\lambda^{j-1}\sigma_j\right)\right]
=\prod_{j=1}^t\cosh\left(z\lambda^{j-1}\right)=\prod_{j=0}^{t-1}\cosh\left(z\lambda^j\right)\,.
\label{pXt-4}
\ee
Thus the non-Markovian random walk with memory has the same probability density function as the Markovian random walk with shrinking steps. 

%%%%%%%%%%%%%%%%%%%%%%%%% fig 3 %%%%%%%%%%%%%%%%%%%%%%%%
\begin{figure} [tbh]
%\vspace{0.2cm}
\epsfxsize=14cm
%\begin{center}
\hskip 16mm
\mbox{\epsfbox{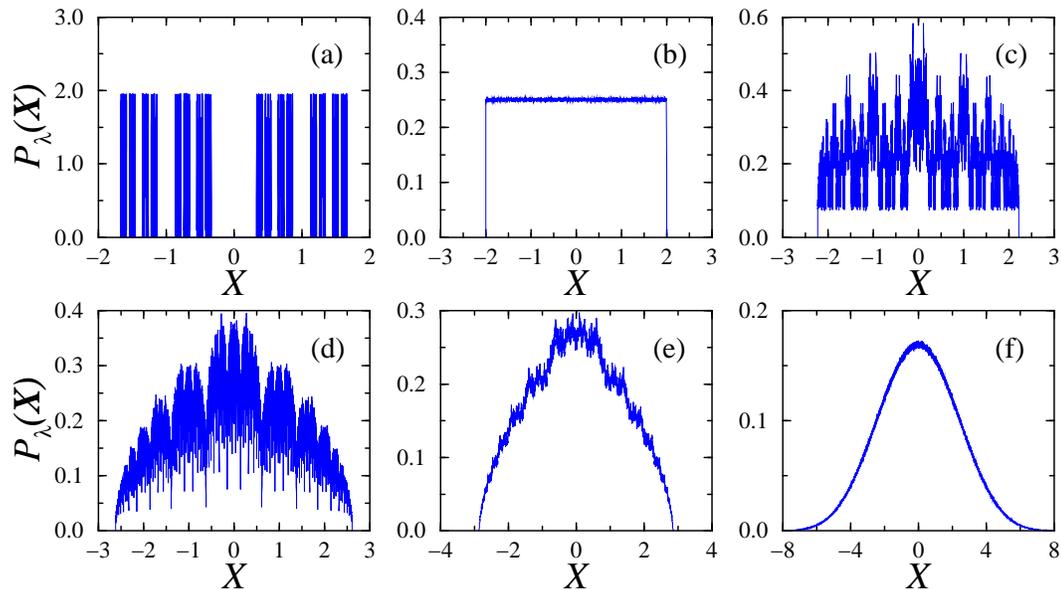}}
%\end{center} 
\vskip -.2cm
\caption{Simulation results for the probability density $P_\lambda(X)$ for $\lambda=0.4$ (a), 0.5 (b), 0.55 (c), $1/\phi=(\sqrt{5}-1)/2=0.618\dots$ (d), 0.65 (e), 0.9 (f). For each value of $\lambda$ the data are collected on $10^8$ samples of the walk, at time $t=50$ to 100 for the largest values of $\lambda$, with a spatial resolution of $10^{-3}$.}
\label{fig3-rw4}  \vskip 0cm
\end{figure}
%%%%%%%%%%%%%%%%%%%%%%%%%%%%%%%%%%%%%%%%%%%%%%%%%%%%

The evolution of the asymptotic behaviour of the probability density as $\lambda$ is varied is illustrated in figure \ref{fig3-rw4}. It evolves from a Cantor set behaviour of the support for small values of $\lambda$, through a uniform density for $\lambda=1/2$, to a Gaussian density when $\lambda\to 1$. The extremes are easily explained by looking at equation~\eref{dp}. When $\lambda\to0$, $X_t\to\sigma_t$ and one obtains the singular Bernoulli density with two delta peaks at $\pm1$. When $\lambda\to1$, $X_t\to\sum_{j=1}^t\sigma_j$ and the density is Gaussian. When $\lambda=1/2$ and $t\to\infty$, the characteristic function is given by
\be
P_{1/2}(ik,\infty)=\prod_{j=0}^{\infty}\cos\left(\frac{k}{2^j}\right)=\frac{\sin(2k)}{2k}\,
\label{cf}
\ee
which is the Fourier transform of the uniform density 
\be
P_{1/2}(X)=\left\{
\begin{array}{lll}
1/4 & , & X\in[-2,2]\\
0 & , & {\rm otherwise}
\end{array}
\right.\,.
\label{ud}
\ee

The probability density $P_\lambda(X)$ is known to be either absolutely continuous or purely singular depending on the values of $\lambda$~\cite{jessen35}. For $\lambda<1/2$ its support is a Cantor set with zero Lebesgue measure so that $P_\lambda(X)$ is singular. It has been shown by Solomyak~\cite{solomyak95} that the cumulative distribution is absolutely continuous for almost all $\lambda>1/2$. But Erd\"os~\cite{erdos39} showed that $P_\lambda(X)$ is singular in $[1/2,1]$ for an infinite set of $\lambda$ values such that $1/\lambda$ is a Pisot number \cite{pisot} like $\phi=(\sqrt{5}+1)/2$, the golden ratio.

\begin{figure} [tbh]
%\vspace{0.2cm}
\epsfxsize=14cm
%\begin{center}
\hskip 16mm
\mbox{\epsfbox{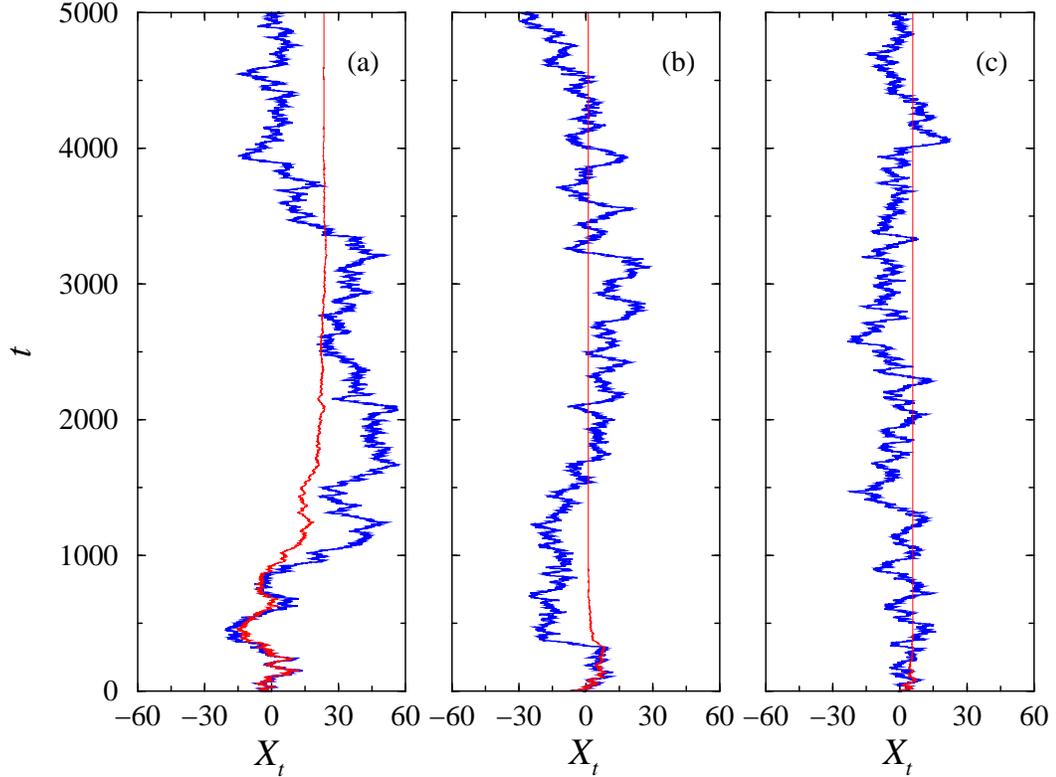}}
%\end{center} 
\vskip -.2cm
\caption{Random walks with memory (thick blue line) and random walks with shrinking steps (thin red line) generated with the same random numbers for $\lambda=0.999$ (a), 0.995 (b) and 0.99 (c). Although these random walks share the same probability density, their two-time statistical properties are quite different.}
\label{fig4-rw4}  \vskip 0cm
\end{figure}

Although the  two walks share the same probability density, their time evolutions are quite different as shown in figure \ref{fig4-rw4} where the random walk with memory keeps on fluctuating at long time whereas the random walk with shrinking steps is quickly frozen.
This difference can be put in evidence looking at the behaviour of the two-time correlation functions. Let us first consider the random walk with shrinking steps. Making use of the property of the two-spin correlation function,
$\langle\sigma_i\sigma_j\rangle=\delta_{i,j}$, equation \eref{def-4} leads to
\be
\langle X_tX_{t+\tau}\rangle=\sum_{i=1}^t\sum_{j=1}^{t+\tau}\lambda^{i+j-2}\langle\sigma_i\sigma_j\rangle
=\sum_{i=1}^t\lambda^{2(i-1)}=\frac{1\!-\lambda^{2t}}{1\!-\lambda^2}=\langle X_t^2\rangle\,,
\label{corr-1}
\ee
%%%%%%%%%%%%%%%%%%%%%%%%% fig 4 %%%%%%%%%%%%%%%%%%%%%%%%
%%%%%%%%%%%%%%%%%%%%%%%%%%%%%%%%%%%%%%%%%%%%%%%%%%%%
whereas one obtains
\be\fl
\langle X_tX_{t+\tau}\rangle=\sum_{i=1}^t\sum_{j=1}^{t+\tau}\lambda^{2t+\tau-i-j}\langle\sigma_i\sigma_j\rangle
=\lambda^\tau\sum_{i=1}^t\lambda^{2(t-i)}=\lambda^\tau\frac{1\!-\lambda^{2t}}{1\!-\lambda^2}=\langle X_t^2\rangle\lambda^\tau
\label{corr-3}
\ee
for the random with memory where the two-time correlation function decays exponentially with $\tau$.
\section{Conclusion}
In this work we have shown how the wandering of a random walker can be efficiently controlled by changing the step lengths at each step. Reducing them in the direction of the previous step while keeping their sum constant leads to a non-diffusive random walk. Surprisingly the probability density is the same as for a random walk with systematically shrinking steps which shows quite different fluctuations (see figure \ref{fig4-rw4}) as shown for the two-point correlation functions.

Instead of controlling the wandering by changing the step lengths, one can modify the jump probabilities to the right and to the left, depending on the direction of the previous step. A preliminary study indicates that in this case the walk remains diffusive \cite{model}.

\end{document}